\documentclass[showpacs,preprintnumbers,nofootinbib,amsmath,amssymb]{revtex4}
\usepackage{hyperref,slashed,color}
\usepackage{graphicx,subfig}
\usepackage{ulem}
\allowdisplaybreaks

\begin{document}
\preprint{TUHEP-TH-13179}
\title{Minimal Ward-Takahashi Vertices and  Pion Light Cone Distribution Amplitudes \\from Gauge Invariant, Nonlocal, Dynamical Quark Model}

\bigskip
\author{Chuan Li$^{2}$\footnote{Email:\href{mailto:lcsyhshy2008@yahoo.com.cn}{lcsyhshy2008@yahoo.com.cn}.},~
Shao-Zhou Jiang$^3$\footnote{Email:\href{mailto:jsz@gxu.edu.cn}{jsz@gxu.edu.cn}.}
 and Qing Wang$^{1,2}$\footnote{Email:
\href{mailto:wangq@mail.tsinghua.edu.cn}{wangq@mail.tsinghua.edu.cn}.}\footnote{corresponding
author}\\~}

\bigskip
\affiliation{$^1$Center for High Energy Physics, Tsinghua University, Beijing 100084, P.R.China\\
$^2$Department of Physics, Tsinghua University, Beijing 100084, P.R.China\footnote{mailing address}\\
$^3$College of Physics Science and Technology, Guangxi University, Nanning, Guangxi 530004, P.R.China}

\begin{abstract}
The gauge-invariant, nonlocal, dynamical quark model is proved to generate the minimal vertices which satisfy the Ward-Takahashi identities. In the chiral limit, the momentum-dependent quark self-energy results in a flat-like form
 with some end point $\delta-$funtions for the light-cone pion distribution amplitudes, similarly found in the Nambu Jona-Lasino model with constant constituent mass.  The leading order nonzero pion and current quark masses corrections lead concave type asymptotic-like form modifications to twist-2 pion distribution amplitude with end point pillars and twist-3 tensor pion distribution amplitude above the flat-like form backgrounds. A by-product of our investigation shows that the variable $u$ appearing in pion light-cone distribution amplitudes is just the standard Feynman parameter in the Feynman parameter integrals; also chiral perturbation works well for these amplitudes.
\end{abstract}

 \maketitle
The gauge-invariant, nonlocal, dynamical quark (GND) model \cite{GND} is one of a phenomenological non-local chiral quark model that can be derived from QCD first principles by a series approximation \cite{WQ2002}. It appears as an approximate description for the effective interactions among light quarks and pseudo-scalar mesons induced from the underlying QCD after integrating out gluon and heavy-quark fields. In the low-energy region, the validity of the GND model is tested by its resulting low-energy constants \cite{GND,WQ2002,WQ2010,WQseries} for the well-known Gasser-Leutwyler chiral Lagrangian \cite{GS}, which well match the existing experimental data for pseudo-scalar meson physics. Considering that in obtaining the GND model the low-energy expansion is not taken, we expect this model has a larger range of application than the traditional low-energy region and are interested in its momentum-dependent behavior beyond the conventional low-energy expansion, which reflects the interplay between low and high energies. In fact, the typical feature of the GND model is its quark self-energy (or momentum-dependent quark mass) $\Sigma(-p^2)$ which represents the original idea of dynamical perturbation theory proposed by Pagels and Stokar \cite{PS}. All dynamics, especially the momentum-dependent effects of the model, are in the main effectively described by this quark self-energy.

To investigate the momentum-dependent effects in terms of a nonlocal model in general, the first obvious problem one faces is its possible violation of the Ward-Takahashi identities (WTIs), as this happens for most nonlocal chiral quark models. In the literature, one way to solve the problem is to artificially revise the vertex \cite{Rvertex,Rvertex1}; another way is to modify the model itself. The GND model belongs to this second approach, where the model is constructed in such a way that it is invariant under local chiral symmetry transformations for external current sources of light-quark fields. Except for the GND model, an earlier GNC model \cite{GNC} took the same tack but with a complex face factor introduced in the model. Because the local chiral symmetry is inserted into the model at inception, we expect the WTIs to hold as a result. This expected validity of the WTIs was claimed but not explicitly shown in the GNC model, and not even mentioned in the original GND model. It is the principle aim of this paper to explicitly show that the GND model does satisfy the WTIs at the chiral limit. A by-product of this demonstration is that we not only give vector and axial-vector vertices, but also scalar, pseudo-scalar, and tensor vertices. In fact, due to their non-perturbative natures, these fundamental vertices have not been very well reported in the literature. The latest ansatz available for the vector vertex in QED is given in Refs.\cite{VectorVertex} which can be traced from the early Refs.\cite{Rvertex}. For vector and axial-vector vertices, the WTIs only constrain their longitudinal parts and cannot fix the transverse parts. For the remaining scalar, pseudo-scalar and tensor vertices, we even have no corresponding WTIs. Considering the vertices we obtained are from the unique GND model action and are constrained by inherent local chiral symmetry of the model, all different types of resulting vertices are at the same level as the approximate description of the corresponding QCD ones. We will show that the vector and axial-vector vertices we obtained in the chiral limit are just the simplest versions satisfying the WTIs and have exactly the same form of those assumed by Ref.\cite{Holdom}; we call these two vertices and related scalar, pseudo-scalar, and tensor vertices the {\it minimal Ward-Takahashi vertices}.

With proof of validity of the WTIs for the GND model and the resulting minimal vertices as our starting point to investigate the momentum-dependent effects, we take the light-cone pion distribution amplitudes (PDAs) as our next target of study in this paper. The reason these are chosen for discussion is that, at present, there is still no definite conclusion on whether PDA is in asymptotic-like form \cite{asymp}, in Chernyak-Zhitnitsky (CZ)-like form \cite{CZ}, or in a flat-like form \cite{flat}. With progress from experiments, ever more information and constraints have emerged on the PDAs providing good crosschecks between theoretical estimations and experimental data \cite{WXG}.

Theoretically, the model computation of PDAs is mainly through chiral quark models. Earlier calculations based on the local chiral quark model or local Nambu Jona-Lasino (NJL)-like models (see Ref.\cite{flat} and references therein) yielded the typical result that the lowest twist-2 PDA is in flat-like form in the chiral limit. In local chiral quark models, quarks have constant constituent masses and there is ultraviolet divergence in the resultant PDAs. To avoid the divergence, various regularization schemes are exploited which often yield confusing results. An improvement is to change to nonlocal chiral quark models (see Ref.\cite{nonlocalPDAs} and references therein), where a momentum-dependent quark mass or quark self-energy is arranged in such a way that it provides the model with a natural soft ultra-violet cutoff and results in finite PDAs. Some researchers believe that the momentum dependence of the quark self-energy will lead to twist-2 PDA deviating from flat-like form and generate correct end-point behavior \cite{nonlocalPDAs1}. As we mentioned previously, a nonlocal chiral quark model usually violates the WTIs and the vertices are revised to avoid the defect. Now, our GND model automatically satisfies the WTIs, so there is no need to make such artificial modification of the vertices.

Another technical problem met in nonlocal chiral quark model is that momentum integrations involving quark self-energies are usually difficult to complete, because usually these have on-shell external momenta which are time-like whereas conventionally loop momenta is space-like because the Wick rotation of the integration momenta occurs in Euclidean space. This mixed appearance of different kinds of momentum variables might cause variable $p^2$ in the quark self-energy $\Sigma(-p^2)$ in the time-like region or even with imaginary components. If there exists an analytical expression of $\Sigma(-p^2)$, such as for instantons \cite{instanton} or just some simple ansatz \cite{Holdom}, there will be no problem, because the extension to the time-like region or imaginary region is explicit. If $\Sigma(-p^2)$ however is defined as the solution of the Schwinger-Dyson equation (SDE), we meet a difficulty. The conventional numerical solution of the SDE is only defined in a space-like region; beyond that, the solution has still not been investigated well \cite{anlycity}. In this paper, chiral perturbation is exploited to overcome this difficulty, with the finding that this perturbation works well.

In the following, we first give a short review of the GND model, next compute vertices, proving that our vertices satisfy the standard WTIs, and finally discuss PDAs. We start with the generating functional of QCD,
\begin{eqnarray}
e^{iW[\overline{I},I,J]}=\int\mathcal{D}\overline{\psi}\mathcal{D}\psi\mathcal{D}\overline{\Psi}\mathcal{D}\Psi\mathcal{D}A_\mu^\alpha~e^{i\int d^4x[-\frac{1}{4}G_{\mu\nu}^{\alpha}G^{\alpha\mu\nu}+\overline{\Psi}(i\slashed{\partial}-g\frac{\lambda^\alpha}{2}\slashed{A}^\alpha-M)\Psi
+\overline{\psi}(i\slashed{\partial}-g\frac{\lambda^\alpha}{2}\slashed{A}^\alpha+J)\psi
+\overline{I}\psi+\overline{I}\psi]}\;,\label{genQCD}
\end{eqnarray}
where $\overline{\psi}$ and $\psi$ are light-quark fields (only u and d quark are taken as light quarks in this paper), $\overline{\Psi}$ and $\Psi$ are heavy-quark fields with mass $M$, $A_\mu^\alpha$ is the gluon field and $G_{\mu\nu}^\alpha$ its field strength. $\overline{I}$ and $I$ are external sources for light-quark fields; $J$ is the external current source for bilinear local light-quark fields. According to its $\gamma$ matrix structure, it can be decomposed into vector, axial-vector, scalar, pseudo-scalar, and tensor parts
\begin{eqnarray}
J(x)=\slashed{v}(x)+\slashed{a}(x)\gamma_5-s(x)+ip(x)\gamma_5+\bar{t}^{\mu\nu}(x)\sigma_{\mu\nu}\;.
\end{eqnarray}
The current quark mass $m$ for light quarks are absorbed into the scalar source $s(x)$; i.e. on the vacuum, all external sources vanish, except $s(x)=m$. After integrating out gluon and heavy-quark fields and integrating in the pseudo-scalar meson field, according to the discussion of Ref.\cite{WQ2002}, the generating functional (\ref{genQCD}) can be approximated by
 \begin{eqnarray}
&&e^{iW[\overline{I},I,J]}=\int\mathcal{D}U\mathcal{D}\overline{\psi}_\Omega\mathcal{D}\psi_\Omega~
e^{iS_{\mathrm{GND}}[\overline{\psi}_\Omega,\psi_\Omega,J_\Omega,\overline{I}_\Omega,I_\Omega]}\label{Gen-1}\\
&&S_{\mathrm{GND}}[\overline{\psi}_\Omega,\psi_\Omega,J_\Omega,\overline{I}_\Omega,I_\Omega]=
\int d^4x\bigg[\overline{\psi}_\Omega[i\slashed{\partial}+J_\Omega-\Sigma(\overline{\nabla}^2)]\psi_\Omega
+\overline{I}_\Omega\psi_\Omega+\overline{I}_\Omega\psi_\Omega\bigg]\label{GNDaction}\\
&&\psi_\Omega=[\Omega^{\dag}P_R+\Omega P_L]\psi\hspace*{0.8cm}\overline{\psi}_\Omega=\overline{\psi}[\Omega^{\dag}P_R+\Omega P_L]\hspace*{1cm}
I_\Omega=[\Omega^{\dag}P_L+\Omega P_R]I\hspace*{0.8cm}\overline{I}_\Omega=\overline{I}[\Omega^{\dag}P_L+\Omega P_R]\\
&&J_\Omega=[\Omega^{\dag}P_L+\Omega P_R][J+i\slashed{\partial}][\Omega^{\dag}P_L+\Omega P_R]=
\slashed{v}_\Omega+\slashed{a}_\Omega\gamma_5-s_\Omega+ip_\Omega\gamma_5+\sigma_{\mu\nu}\bar{t}^{\mu\nu}_\Omega\\
&&U=\Omega^2\hspace*{2cm}\overline{\nabla}^\mu=\partial^\mu-iv_\Omega^\mu\;,
\end{eqnarray}
where $U$ and $\Omega$ are unimodular pseudo-scalar meson fields, $\overline{\psi}_\Omega$,$\psi_\Omega$ and $\overline{I}_\Omega$,$I_\Omega$ are rotated light-quark fields and corresponding sources, respectively, $J_\Omega$ is the rotated source for the rotated bilinear quark currents. $\Sigma(-p^2)$ is the self-energy for the rotated light-quark fields which satisfy the corresponding SDE. $S_{\mathrm{GND}}[\overline{\psi}_\Omega,\psi_\Omega,J_\Omega,\overline{I}_\Omega,I_\Omega]$ is the action of the GND model; in the original paper \cite{GND}, $\overline{I}$ and $I$ were not introduced, but an extra normalization term $i\mathrm{Trln}[i\slashed{\partial}+J_\Omega]$ appeared. In Ref.\cite{WQ2010}, this extra term was later proven to drop out. Integrating out the rotated light-quark and pseudo-scalar meson fields of (\ref{Gen-1}), we obtain
\begin{eqnarray}
&&W[\overline{I},I,J]=-i\mathrm{Trln}D^{-1}-\int d^4xd^4y\overline{I}_\Omega(x)D(x,y)I_\Omega(y)+\mbox{pseudo-scalar meson loop corrections}\label{StartGND}\\
&&D^{-1}(x,y)=[i\slashed{\partial}_x+J_\Omega-\Sigma(\overline{\nabla}_x^2)]\delta(x-y)\;,
\end{eqnarray}
where the pseudo-scalar field $\Omega$ in (\ref{StartGND}) must satisfy the stationary equation $\frac{\partial\mathrm{Trln}D^{-1}}{\partial\Omega(x)}=0$ for $\overline{I}=I=0$. Considering that in the low-energy region, we have shown in \cite{WQ2002} that the Lagrangian of (\ref{StartGND}) just presumes the standard Gasser-Leutwyler chiral Lagrangian \cite{GS}, the stationary equation then can be replaced with the equation of motion (EOM) derived from the Gasser-Leutwyler chiral Lagrangian. For convenience in computation, we use below this chiral Lagrangian-induced EOM for $\Omega$, because it is more common and relatively simple.\vspace*{1cm}~

Neglecting pseudo-scalar meson loop corrections, the vertices in the chiral limit are defined by the following 3-point Green's function
\begin{eqnarray}
&&\hspace*{-0.5cm}\langle 0|\mathbf{T}\psi(x)\overline{\psi}(y)\overline{\psi}(z)\tilde{\gamma}_i\tau^a\psi(z)|0\rangle_{m=0}=-\frac{\delta^3 W[\overline{I},I,J]}{\delta\overline{I}(x)\delta I(y)\delta J_i^a(z)}\bigg|_{J=0,\overline{I}=I=0}
\equiv\int d^4x'd^4y'D(x,x')\Gamma_i^a(x',y',z)D(y'y)\\
&&\hspace{-0.5cm}=-\frac{\delta}{\delta J_i^a(z)}\bigg[[\Omega^\dag(x)P_L+\Omega(x)P_R]D(x,y)[\Omega^\dag(y)P_L+\Omega(y)P_R]\bigg]\bigg|_{J=0,\overline{I}=I=0}\;,
\nonumber
\end{eqnarray}
which yields
\begin{eqnarray}
\Gamma_i^a(x,y,z)
=\bigg[\frac{\delta D^{-1}(x,y)}{\delta J_i^a(z)}-D^{-1}(x,y)\frac{\delta\Omega(y)}{\delta J_i^a(z)}\gamma_5-\frac{\delta\Omega(x)}{\delta J_i^a(z)}\gamma_5D^{-1}(x,y)\bigg]\bigg|_{J=0,\overline{I}=I=0}\;,
\end{eqnarray}
where $i=S,P,V,A,T$, $\tau^a$ are the Pauli matrices in isospin space with $\tau^0=1$, and
\begin{eqnarray}
&&\hspace*{-1cm}J_S^a=-s^a\hspace*{0.3cm}J_P^a=p^a\hspace*{0.3cm}
J_V^a=v_\mu^a\hspace*{0.3cm}J_A^a=a_\mu^a\hspace*{0.3cm}J_T^a=\bar{t}_{\mu\nu}^a
\hspace*{1cm}\tilde{\gamma}_S=1\hspace*{0.3cm}\tilde{\gamma}_P=i\gamma_5\hspace*{0.3cm}
\tilde{\gamma}_V=\gamma^\mu\hspace*{0.3cm}\tilde{\gamma}_A=\gamma^\mu\gamma_5\hspace*{0.3cm}\tilde{\gamma}_T=\sigma^{\mu\nu}\nonumber
\end{eqnarray}
with the help of the following EOM results in the chiral limit
\begin{eqnarray}
&&\frac{\delta\Omega(x)}{\delta s^a(y)}\bigg|_{J=0,\overline{I}=I=0}=\frac{\delta\Omega(x)}{\delta v_\mu^a(y)}\bigg|_{J=0,\overline{I}=I=0}=\frac{\delta\Omega(x)}{\delta \bar{t}_{\mu\nu}^a(y)}\bigg|_{J=0,\overline{I}=I=0}=0\label{Omegaderrivative1}\\
&&\frac{\delta\Omega(x)}{\delta p^a(y)}\bigg|_{J=0,\overline{I}=I=0}=\tau^a(1-\delta_{a0})\frac{iB_0}{\partial^2_x}\delta(x-y)\hspace*{1cm}
\frac{\delta\Omega(x)}{\delta a_\mu^a(y)}\bigg|_{J=0,\overline{I}=I=0}=\bigg[\tau^a(1-\delta_{a0})+\delta_{a0}\bigg]\frac{i\partial^\mu_x}{\partial^2_x}\delta(x-y)\label{Omegaderrivative2}\;,
\end{eqnarray}
where $B_0=-\frac{1}{2}\langle\overline{\psi}\psi\rangle/F_0^2$ is the $p^2$-order low-energy constant of the Gasser-Leutwyler chiral Lagrangian related to the ratio of the quark condensate $\langle\overline{\psi}\psi\rangle$ and the square of the pion decay constant, $F_0^2$. With the help of (\ref{Omegaderrivative1}) and (\ref{Omegaderrivative2}), we can compute a series derivation of rotated sources $s_\Omega,p_\Omega,v_\Omega,a_\Omega,\bar{t}_\Omega$ to un-rotated sources $s,p,v,a,\bar{t}$. With these relations, we obtain the chiral-limit result
\begin{eqnarray}
&&\hspace{-0.5cm}\Gamma_S^a(x,y,z)=\tau^a\delta(x-z)\delta(y-z)\hspace*{2cm}\Gamma_{T,\mu\nu}^a(x,y,z)=\tau^a\sigma_{\mu\nu}\delta(x-z)\delta(y-z)
\label{SandT}\\
&&\hspace{-0.5cm}\Gamma_i^a(x,y,z)=\int\frac{d^4pd^4q}{(2\pi)^8}e^{-iq\cdot x+ip\cdot y+i(q-p)\cdot z}\tilde{\Gamma}_i^a(p,q)\hspace*{1cm}i=P,V,A\;.
\end{eqnarray}
Whereas the resulting scalar and tensor vertices are trivial, the pseudo-scalar, vector, and axial-vector vertices are non-trivial, their momentum space expressions being
\begin{eqnarray}
&&\tilde{\Gamma}_P^a(p,q)=i\tau^a\bigg[1-B_0(1-\delta_{a0})\frac{\Sigma(-q^2)+\Sigma(-p^2)}{(q-p)^2}\bigg]\gamma_5\label{pseuscalar}\\
&&\tilde{\Gamma}_{V,\mu}^a(p,q)=\tau^a\bigg[\gamma_\mu-\frac{q_\mu+p_\mu}{q^2-p^2}[\Sigma(-q^2)-\Sigma(-p^2)]\bigg]\label{vector}\\
&&\tilde{\Gamma}_{A,\mu}^a(p,q)=\tau^a\bigg[\gamma_\mu-\frac{q_\mu-p_\mu}{(q-p)^2}[\Sigma(-q^2)+\Sigma(-p^2)]\bigg]\gamma_5\;.\label{axialvector}
\end{eqnarray}
Equations (\ref{vector}) and (\ref{axialvector}) have exactly the same forms as those assumed by Ref.\cite{Holdom}.
From (\ref{vector}), the WTI for the vector vertex is
\begin{eqnarray}
(q-p)^\mu\tilde{\Gamma}_{V,\mu}^a=\tau^a[S^{-1}(q)-S^{-1}(p)]\hspace*{2cm}S^{-1}(p)=\slashed{p}-\Sigma(-p^2)\;,
\end{eqnarray}
where $S(p)$ is the light-quark propagator in momentum space with  $D(x,y)\bigg|_{J=0}=\int\frac{d^4p}{(2\pi)^4}e^{-ip\cdot x}S(p)$. From (\ref{axialvector}), the WTI for the axial-vector vertex is
\begin{eqnarray}
(q-p)^\mu\tilde{\Gamma}_{A,\mu}^a=\tau^a[S^{-1}(q)\gamma_5+\gamma_5S^{-1}(p)]\;.
\end{eqnarray}
The above two WTIs are standard and yield the minimal solutions for (\ref{vector}) and (\ref{axialvector}). We call (\ref{SandT}), (\ref{pseuscalar}), (\ref{vector}) and (\ref{axialvector}) the minimal WT vertices and the corresponding GND model is the model which correctly generates these minimal WT vertices. In other words, if one prefers to trace the minimal WT vertices back to a source effective action, it is just the GND model action (\ref{GNDaction}).

Considering that the present GND model is only an approximation of the underlying QCD, we will in the future discuss its corrections in QCD. That will mean amending the above minimal WT vertices.\vspace*{0.5cm}~

We shall now discuss the light-cone PDAs, their definitions being as in Ref.\cite{DAdef}
\begin{eqnarray}
\langle\vec{p}|\psi(x)\overline{\psi}(0)|0\rangle&=&-\frac{if}{4}\int_0^1du~e^{i(1-u)p\cdot x}~\Phi(u,p,x)\;,\label{PhiDef}\\
\Phi(u,p,x)&=&\bigg[\slashed{p}\gamma_5\phi(u)
-\frac{m_\pi^2}{2m}\gamma_5[\phi_p(u)+\sigma_{\mu\nu}p^{\mu}x^\nu\frac{\phi_\sigma(u)}{6}]+\mbox{high twist corrections}\bigg]_{p^0=\sqrt{\vec{p}^2+m_\pi^2}}\;,\label{PDAs}
\end{eqnarray}
where $\langle\vec{p}|$ is a pion state with momentum $\vec{p}$, $m_\pi,m_u,m_d$ are masses of the pion, u quark, and d quark respectively. For simplicity, we ignore the mass difference between u and d quarks setting them to the same current mass $m=m_u=m_d$. $\phi(u)$ is the leading twist or twist-2 PDA, $\psi_p(u)$ and $\phi_\sigma(u)$ are sub-leading or twist-3 PDAs that correspond to the pseudo-scalar and the pseudo-tensor structures respectively. In our model, the operator $\psi(x)\overline{\psi}(0)$ appeared in (\ref{PhiDef}) is related to $\frac{\delta^2W[\overline{I},I,J]}{\delta\overline{I}(x)\delta I(0)}$, and its contribution to the pion matrix element relies on the part proportional to the pion field in the result. Parameterizing the pion field $\Pi$ by $\Omega =e^{i\Pi/2}$, the rotated source becomes
\begin{eqnarray}
J_\Omega\bigg|_{J=-m,\overline{I}=I=0}
=[1+i\frac{\Pi}{2}\gamma_5][-m+i\slashed{\partial}][1+i\frac{\Pi}{2}\gamma_5]=
-m-(im\Pi+\frac{1}{2}\slashed{\partial}\Pi)\gamma_5+O(\Pi^2)\;.
\end{eqnarray}
Hence,
\begin{eqnarray}
&&\hspace{-0.5cm}\langle\vec{p}|\psi(x)\overline{\psi}(0)|0\rangle=\langle\vec{p}|\frac{\delta^2W[\overline{I},I,J]}{\delta\overline{I}(x)\delta I(0)}|0\rangle\bigg|_{J=-m,\overline{I}=I=0}\nonumber\\
&&\hspace{-0.5cm}=\langle\vec{p}|[1+\frac{i}{2}\Pi(x)\gamma_5]\bigg[\frac{1}{i\slashed{\partial}-m-\Sigma(\partial^2)
-(im\Pi+\frac{1}{2}\slashed{\partial}\Pi)\gamma_5}\bigg](x,0)[1+\frac{i}{2}\Pi(x)\gamma_5]|0\rangle
\nonumber\\
&&\hspace{-0.5cm}=i\langle\vec{p}|\Pi(0)|0\rangle\int\frac{d^4q}{(2\pi)^4}e^{iq\cdot x}\bigg[\gamma_5e^{ip\cdot x}\frac{m+\Sigma(-q^2)}{q^2-[m+\Sigma(-q^2)]^2}-\frac{1}{\slashed{q}\!+\!m\!+\!\Sigma(-q^2)}
(m+\frac{\slashed{p}}{2})\gamma_5\frac{1}{\slashed{p}\!-\!\slashed{q}\!-\!m\!-\!\Sigma[-(p-q)^2]}\bigg]\;,
\end{eqnarray}
where the first term, after expanding $e^{iq\cdot x}$ in terms of powers of $q\cdot x$ and ignoring high twist $O(x^2)$ terms, contributes to the twist-3 pseudo-scalar PDA $\phi_p(u)$ at the end-point $u=0$, whereas the second term can be changed to the form of (\ref{PhiDef}) by the standard Feynman parametrization method,
\begin{eqnarray}
&&\hspace{-1cm}\int\frac{d^4q}{(2\pi)^4}e^{iq\cdot x}\frac{1}{\slashed{q}\!+\!\tilde{\Sigma}(q)}
(m+\frac{\slashed{p}}{2})\gamma_5\frac{1}{\slashed{p}\!-\!\slashed{q}\!-\!\tilde{\Sigma}(p-q)}=
\int\frac{d^4q}{(2\pi)^4}e^{iq\cdot x}\frac{\slashed{q}\!-\!\tilde{\Sigma}(q)}{q^2\!-\!\tilde{\Sigma}^2(q)}
(m+\frac{\slashed{p}}{2})\gamma_5\frac{\slashed{p}\!-\!\slashed{q}\!+\!\tilde{\Sigma}(p-q)}{(p-q)^2\!-\!\tilde{\Sigma}^2(p-q)}
\nonumber\\
&&\hspace{-1cm}=\int_0^1du~e^{i(1-u)p{\cdot}x}\int\frac{d^4q}{(2\pi)^4}e^{iq{\cdot}x}
\frac{[\slashed{q}+(1-u)\slashed{p}\!-\!\tilde{\Sigma}(q+p-up)](m+\frac{\slashed{p}}{2})
[-u\slashed{p}\!+\!\slashed{q}\!+\!\tilde{\Sigma}(up-q)]}{\{q^2+u(1-u)p^2-\tilde{\Sigma}^2(q+p-up)
+(1-u)[\tilde{\Sigma}^2(q+p-up)-\tilde{\Sigma}^2(up-q)]\}^2}\gamma_5\;,\label{MomInt}
\end{eqnarray}
where $u$ is the standard Feynman parameter appearing in the well-known Feynman parameter integration and $\tilde{\Sigma}(q)$ is the abbreviated expression for $m\!+\!\Sigma(-q^2)$. Conventionally, the Feynman parameter integration is used to deal with loop-momentum integration with constant mass in the perturbation theory; here, applied to the momentum integration involving momentum-dependent quark self-energy, it still works. One can easily check that if we take the quark self-energy appearing in (\ref{MomInt}) back to constant mass as in the traditional NJL model, (\ref{MomInt}) just becomes the standard Feynman parameterization formula. Indeed, this identification of the Feynman parameter with the PDA variable $u$ not only endows the traditional mathematical Feynman parameter with a physical meaning, but also is valid in any kind of chiral quark or NJL-like models, as long as we have momentum integration of form (\ref{MomInt}) with two quark propagators (the form of vertex between two propagators is not important for this issue), no matter whether the models are local or non-local. This result is independent of the GND model investigated in this paper; the GND model here is taken as an example to exhibit details of the computation .

With the above Feynman parameterization, and after lengthy computations, we finally obtain $\Phi(u,p,x)$ introduced in (\ref{PhiDef}) as
\begin{eqnarray}
-\frac{if}{4}\Phi(u,p,x)&=&i\langle\vec{p}|\Pi(0)|0\rangle\bigg\{\delta(u\!-\!0^+)\int\frac{d^4q}{(2\pi)^4}\gamma_5\frac{\tilde{\Sigma}(q)}{q^2-\tilde{\Sigma}^2(q)}
\nonumber\\
&&-\int\frac{d^4q}{(2\pi)^4}
\frac{I_3\!+\!\frac{p{\cdot}q}{p^2}I_1\!+\!
\left(\frac{q^2}{3}\!-\!\frac{(p{\cdot}q)^2}{3p^2}\right)iI_2}{\{q^2+u(1-u)p^2-\tilde{\Sigma}^2(up-q)
+u[\tilde{\Sigma}^2(up-q)-\tilde{\Sigma}^2(q+p-up)]\}^2}\gamma_5\nonumber\\
&&\hspace*{-2.5cm}-\left[-\delta(u\!-\!1^-)\!+\!\delta(u\!-\!0^+)\!+\!\frac{\partial}{\partial u}\right]\int\frac{d^4q}{(2\pi)^4}
\frac{\left(\frac{q^2}{3}\!-\!\frac{(p{\cdot}q)^2}{3p^2}\right)I_4\!+\!\frac{p{\cdot}q}{p^2}I_3
+\left(-\frac{q^2}{3}\!+\!\frac{4(p{\cdot}q)^2}{3p^2}\right)\frac{I_1}{p^2}}{\{q^2+u(1-u)p^2-\tilde{\Sigma}^2(up-q)
+u[\tilde{\Sigma}^2(up-q)-\tilde{\Sigma}^2(q+p-up)]\}^2}\gamma_5\bigg\}\nonumber\\
&&+\mbox{high twist terms}\;,\label{PhiResult}
\end{eqnarray}
where $p^2=m^2_\pi$ and
\begin{eqnarray}
I_1&=&\slashed{p}[q{\cdot}p+\frac{1}{2}(1-2u)p^2+m\tilde{\Sigma}(up-q)-m\tilde{\Sigma}(q+p-up)]
+\frac{1}{2}p^2[2(1-2u)m+\tilde{\Sigma}(up-q)-\tilde{\Sigma}(q+p-up)]\;,\\
&&\hspace*{-1cm}I_2=\frac{1}{2}x_{\mu}p_{\nu}[\gamma^{\mu},\gamma^{\nu}][\frac{1}{2}\tilde{\Sigma}(up-q)-m\!+\!\frac{1}{2}\tilde{\Sigma}(q+p-up)]\;,~~~\\
&&\hspace{-1cm}I_3=(m-\slashed{p})q^2+[(1-u)\slashed{p}\!-\!\tilde{\Sigma}(q+p-up)](m+\frac{\slashed{p}}{2})[-u\slashed{p}\!+\!\tilde{\Sigma}(up-q)]\;,\\
&&\hspace*{-1cm}I_4=\frac{1}{2}\tilde{\Sigma}(up-q)+(1-2u)m\!-\!\frac{1}{2}\tilde{\Sigma}(q+p-up)\;.
\end{eqnarray}
In (\ref{PhiResult}), we have dropped terms proportional to $\slashed{x}$, since they belong to twist-4 terms. The  delta function terms are just end-point terms that are from non-exponential $p{\cdot}x$ terms appearing in the original momentum integration (\ref{MomInt}) when we expand $e^{iq\cdot x}$ in terms of powers of $q\cdot x$ and apply Lorentz invariance in decomposing its Lorentz structure.

Since (\ref{PhiResult}) already has structure of (\ref{PDAs}), through comparison between the two equations, we can easily read out general expressions of PDAs $\phi(u)$, $\psi_p(u)$ and $\phi_\sigma(u)$
in terms of quark self energy. One can check that except the pure end point term in the first line of (\ref{PhiResult}), all other terms satisfying symmetry $u\leftrightarrow1-u$, since under combined transformation of $u\leftrightarrow1-u$ and $q\leftrightarrow-q$, the integrand of momentum integration is even  for the second line and odd for the third line, respectively. This implies the corresponding symmetry of $u\leftrightarrow1-u$ for result PDAs. In the chiral limit, $m=0$ and $p^2=m^2_\pi=0$, we find (\ref{PhiResult}) simplifies to
\begin{eqnarray}
\Phi(u,p,x)
\stackrel{\mbox{\tiny chiral limit}}{======}&&\hspace*{-0.2cm}\Phi_0(u,p,x)
=-\frac{4\langle\vec{p}|\Pi(0)|0\rangle}{f}\bigg[\slashed{p}\gamma_5\bigg([\delta(u-1^-)+\delta(u-0^+)]\phi_{\delta0}+\phi_0\bigg)\nonumber\\
&&\hspace*{3.9cm}+\delta(u-0^+)\phi_{p,\delta0}
 -\frac{m_\pi^2}{2m}\gamma_5\sigma_{\mu\nu}p^{\mu}x^\nu\frac{\phi_{\sigma,0}}{6}\bigg]\;,
~~~\label{PhiClimitResult}
\end{eqnarray}
where the four coefficients $\phi_{\delta0}$,$\phi_0$,$\phi_{p,\delta0}$ and $\phi_{\sigma,0}$ are expressed as Euclidean-space momentum integrations
\begin{eqnarray}
&&\hspace*{-1cm}
\phi_{\delta0}=-\int\frac{q_E^2 dq_E^2}{16\pi^2}\frac{1}{4}X^2q_E^2\Sigma\Sigma'\hspace*{1.3cm}
\phi_0=\int\frac{q_E^2 dq_E^2}{16\pi^2}X^2[\frac{1}{4}q_E^2-\frac{1}{2}\Sigma^2+\frac{1}{2}q_E^2\Sigma\Sigma']\\
&&\hspace*{-1cm}\phi_{p,\delta0}=\int\frac{q_E^2 dq_E^2}{16\pi^2}X\Sigma\hspace*{2.5cm}
\phi_{\sigma,0}=\frac{3m}{m_\pi^2}\int\frac{q_E^2 dq_E^2}{16\pi^2}X^2q_E^2\Sigma\;,\\
&&\hspace*{-1cm}\Sigma=\Sigma(q_E^2)\hspace*{1cm}\Sigma'=\frac{d\Sigma(q_E^2)}{dq_E^2}\hspace*{1cm}
X=\frac{1}{q_E^2+\Sigma^2(q_E^2)}\;.
\end{eqnarray}
Comparing the above result with (\ref{PDAs}), we find that with the exception of end-point values for $\phi(u)$ symmetrically at u=0 and u=1 and $\phi_p(u)$ non-symmetrically at u=0, PDAs are all independent of $u$ ($\phi_p(u)$ even vanishes) and therefore take flat-like forms. Graphically, the pictures are that two symmetric infinitesimal narrow pillars at two end points appear above the flat-like form backgrounds for $\phi(u)$,  one pure non-symmetric pillar at $u=0$ for $\phi_p(u)$ with zero background, and no  pillar for $\phi_\sigma(u)$ with just flat-like form background. This result is completely due to the Feynman parameter description of PDAs, and therefore is valid for any type of chiral quark or NJL-like models, whether local or non-local (one can check by
replacing quark self energy with constant mass that the end-point terms from $\delta-$function for $\phi(u)$ and $\phi_p(u)$ exist even in local situations). This result is differs from those previously obtained in the literature, where researchers believed that the momentum dependence of the quark self-energy would force PDAs to deviate from flat-like forms in the chiral limit. Considering our result is analytical that does not rely on technical details of numerical computations and is valid for a large class of chiral quark models, we believe it is reliable. This result implies that, at least in the chiral limit, the momentum-dependent behavior of the quark self-energy, or more fundamentally the related non-locality of its interaction, causes no difference in the form of PDAs with constant mass or local interaction such as the NJL model.

For the chiral limit result (\ref{PhiClimitResult}), $\phi(u)$ and $\phi_\sigma(u)$ are ultraviolet divergent. The reason causing this divergence is that the quark self-energy of the GND model is for the rotated quark field. If the quark self-energy instead is for the un-rotated quark fields, as the conventional instanton model does, we need to replace the current quark mass $m$ appearing in the numerator of the l.h.s. term of (\ref{MomInt}) with the quark self-energy and ignore the $\slashed{p}$ term in the same numerator. This then will suppress the ultraviolet behavior of the momentum integration decreasing the ultraviolet divergences of PDAs .

If we want to go beyond the chiral limit, the momentum integrations in the second and third line of (\ref{PhiResult}) are  not easy to achieve. To finish the momentum integrations, we are used to rotate the momentum integration variable $q$ into Euclidean space, although the external momentum $p$ must be kept on the pion mass shell, $p^2=m^2_\pi$. This will create some imaginary components; for example $\Sigma[-(q-up)^2]=\Sigma[-q^2-u^2p^2+2uq{\cdot}p]$ will become $\Sigma[q_E^2-u^2p^2+2u(iq_E^0p^0-\vec{q}_E\cdot\vec{p})]$ after a Wick rotation for integration variable $q^{\mu}=(q^0,\vec{q})\rightarrow(iq_E^0,\vec{q}_E)$. Because the SDE now cannot provide us with a reliable quark self-energy beyond the space-like momentum region, we are not able then to directly compute the momentum integration in (\ref{PhiResult}).
One way to avoid this difficulty is to go back to the original constant quark mass case (or equivalently the NJL model situation), where all momentum integration can be analytically finished except we need some momentum cutoff to regularize the integration. An alternative approach is to set an analytical expression for $\Sigma(-q^2)$, as for instantons \cite{instanton} or introduce some simple ansatz \cite{Holdom}, then the momentum integration can still be finished, at least at the level of numerical computations. Considering that for the former information of the momentum dependence of the quark self-energy is lost, and the latter is constrained by specific choices of quark self-energy which also might not precisely describe its QCD behavior, we propose in this paper to expand the integrand in terms of powers of $p^2=m^2_\pi$ and $m$. The underlying basis for this expansion is that when we go to higher-order terms in the expansion, according to (\ref{PhiResult}), typically we encounter a factor of $[p{\cdot}q/(q^2+\Sigma^2)]^2$ or $p^2/(q^2+\Sigma^2)$ or $mp{\cdot}q/(q^2+\Sigma^2)$. In units of the QCD scale parameter $\Lambda_{\mathrm{QCD}}$, the largest contribution comes mainly from the platform region of the quark self-energy in which $\Sigma/\Lambda_{\mathrm{QCD}}\sim 2$ and $q_E/\Lambda_{\mathrm{QCD}}\leq 1$ because quark self-energies above $\Lambda_{\mathrm{QCD}}$ decrease with momentum as $1/q^2$ \cite{WQ2002} and contribute little. Considering that in our case $\Lambda_{\mathrm{QCD}}\sim 440$MeV is much larger than the pion mass, i.e. $p/\Lambda_{\mathrm{QCD}}\sim m_\pi/\Lambda_{\mathrm{QCD}}\sim1/3$, and $m<10$MeV, then the first factor $[p{\cdot}q/(q^2+\Sigma^2)]^2\sim 10^{-2}$, the second factor $p^2/(q^2+\Sigma^2)\sim 10^{-1}$, and the third factor $mp{\cdot}q/(q^2+\Sigma^2)\sim 10^{-3}$; that is, each factor is at least an order of magnitude small. We expect chiral expansion will works well. With this analysis, the result of the detail computation gives up to first order,
\begin{eqnarray}
\Phi(u,p,x)&=&\Phi_0(u,p,x)+\Phi_1(u,p,x)+O(m^4_\pi,m^2,mm^2_\pi)\;,\label{Up1orderResult}\\
\Phi_1(u,p,x)&=&-\frac{4\langle\vec{p}|\Pi(0)|0\rangle}{f}\bigg[
\slashed{p}\gamma_5\bigg([\delta(u-1^-)+\delta(u-0^+)]\phi_{\delta1}+\phi_{10}+\phi_{11}u(1-u)\bigg)\nonumber\\
&&+\delta(u-0^+)\phi_{p,\delta10}+[\delta(u-1^-)+\delta(u-0^+)]\phi_{p,\delta11}
+\phi_{p1}
 -\frac{m_\pi^2}{2m}\gamma_5\sigma_{\mu\nu}p^{\mu}x^\nu\frac{\phi_{\sigma,10}+\phi_{\sigma,11}u(1-u)}{6}\bigg]
 ~~~\;,\label{Phi1}
\end{eqnarray}
where the eight coefficients $\phi_{\delta1}$,$\phi_{10}$,$\phi_{11}$,$\phi_{p,\delta10}$,$\phi_{p,\delta11}$,$\phi_{p1}$,$\phi_{\sigma,10}$, and $\phi_{\sigma,11}$ are expressed as Euclidean-space momentum integrations
\begin{eqnarray}
\phi_{\delta1}&=&m_\pi^2\int\frac{q_E^2 dq_E^2}{16\pi^2}\bigg[X^2(\frac{1}{4}q_E^2\Sigma\Sigma''
 +\frac{1}{12}q_E^4\Sigma\Sigma''-\frac{1}{6}q_E^4\frac{m}{m_\pi^2}\Sigma'')\bigg]\\
\phi_{10}&=&m_\pi^2\int\frac{q_E^2 dq_E^2}{16\pi^2}\bigg[X^2[(\Sigma-\frac{1}{2}q_E^2\Sigma'+\frac{1}{3}q_E^4\Sigma'')\frac{m}{m_\pi^2}
 +\frac{1}{2} \Sigma\Sigma'-\frac{1}{2}q_E^2\Sigma\Sigma''-\frac{1}{4}q_E^2{\Sigma'}^2\\
 &&-\frac{1}{4}q_E^4\Sigma'\Sigma''-\frac{1}{4}q_E^4\Sigma\Sigma''')
 +X^3(-\frac{1}{2}q_E^2\Sigma\Sigma'-2q_E^2\Sigma^2{\Sigma'}^2-q_E^2\Sigma^3\Sigma''- q_E^4\Sigma{\Sigma'}^3
 -2q_E^4\Sigma^2\Sigma'\Sigma''-\frac{1}{3}q_E^4\Sigma^3\Sigma''')\bigg]\;,\notag\\
 \phi_{11}&=&m_\pi^2\int\frac{q_E^2 dq_E^2}{16\pi^2}\bigg[X^2(-\frac{1}{2}
  -\Sigma\Sigma'+q_E^2{\Sigma'}^2 + q_E^2\Sigma\Sigma''+\frac{3}{2}q_E^4\Sigma'\Sigma'' +\frac{1}{2}q_E^4\Sigma\Sigma''')\\
 &&+X^3( -\Sigma^2 -2 \Sigma^3\Sigma' +\frac{1}{2} q_E^2 +4 q_E^2\Sigma\Sigma' +11q_E^2\Sigma^2{\Sigma'}^2
 +5 q_E^2\Sigma^3\Sigma'' +6 q_E^4\Sigma{\Sigma'}^3 +12 q_E^4\Sigma^2\Sigma'\Sigma''
 +2 q_E^4 \Sigma^3\Sigma''') \bigg]\;,\notag\\
 \phi_{p,\delta10}&=&\int\frac{q_E^2 dq_E^2}{16\pi^2}mX[1-2X\Sigma^2]\;,\\
  \phi_{p,\delta11}&=&2m\int\frac{q_E^2 dq_E^2}{16\pi^2}X^2\bigg[ (-\frac{1}{4}q_E^2
 -\frac{1}{2}q_E^2 \Sigma\Sigma')\frac{m}{m_\pi^2}-\frac{1}{8} q_E^2\Sigma'-\frac{1}{8}q_E^4 \Sigma''\bigg]\;,\\
 \phi_{p1}&=&2m\int\frac{q_E^2 dq_E^2}{16\pi^2}X^2\bigg[ (\Sigma^2 +\frac{3}{2}q_E^2+ q_E^2\Sigma\Sigma')\frac{m}{m_\pi^2} -\frac{1}{2} \Sigma+\frac{1}{4}q_E^4\Sigma''\bigg]\;,\\
 \phi_{\sigma,10}&=&12m\int\frac{q_E^2 dq_E^2}{16\pi^2}X^2\bigg[
  -\frac{1}{4}q_E^2\frac{m}{m_\pi^2} -\frac{1}{8}q_E^2\Sigma' -\frac{1}{24}q_E^4\Sigma''\bigg]\;,\\
    \phi_{\sigma,11}&=&12m\int\frac{q_E^2 dq_E^2}{16\pi^2}X^2\bigg[\frac{1}{4}q_E^2\Sigma'+\frac{1}{12}q_E^4\Sigma''\bigg]\;,\\
   &&\hspace*{-1cm}\Sigma''=\frac{d^2\Sigma(q_E^2)}{d(q_E^2)^2}\hspace*{1cm}
 \Sigma'''=\frac{d^3\Sigma(q_E^2)}{d(q_E^2)^3}\;.
 \end{eqnarray}
We see that the first-order corrections include three parts
\begin{itemize}
\item Corrections to chiral limit result which include: heights to two symmetric infinitesimal narrow pillars at two end points from $\phi_{\delta1}$ and  to the flat-like form backgrounds from $\phi_{10}$ for $\phi(u)$;
heights to non-symmetric pillar at $u=0$ from $\phi_{p,\delta10}$ and to zero backgrounds from $\phi_{p1}$ for $\phi_p(u)$; and height for flat-like form background from $\phi_{\sigma,10}$ for $\phi_\sigma(u)$.
\item Two symmetric infinitesimal narrow negative pillars at two end points from $\phi_{p,\delta11}$ for $\phi_p(u)$ appear.
\item Asymptotic-like form concave type corrections \cite{asymp} proportional to $u(1-u)$ from $\phi_{11}$ and $\phi_{\sigma,11}$ to $\phi(u)$ and  $\phi_\sigma(u)$ respectively.
\end{itemize}
To compare with literature's CZ-like form results, traditional double-hump structure of PDAs now for $\phi(u)$ is squeezed to two symmetric infinitesimal narrow pillars at two end points plus concave type asymptotic-like form above flat-like form background; for $\phi_p(u)$ is squeezed to two non-symmetric infinitesimal narrow pillars at two end points above a flat-like form background; for $\phi_\sigma(u)$ is changed to concave type asymptotic-like form above flat-like form background. We expect more complex  non-asymptotic-like form corrections will show up in more higher order of our chiral perturbation expansion.

If we further consider the normalization condition for $\phi(u)$,
\begin{eqnarray}
\int_0^1du~\phi(u)=1\;.
\end{eqnarray}
then our results (\ref{PhiClimitResult}), (\ref{Up1orderResult}) and (\ref{Phi1}) imply that
\begin{eqnarray}
-4\langle\vec{p}|\Pi(0)|0\rangle(2\phi_{\delta0}+2\phi_{\delta1}+\phi_0+\phi_{10}+\frac{1}{6}\phi_{11})=1\;.
\end{eqnarray}
To obtain numerical results, we solve the SDE obtaining a quark self-energy as in Ref.\cite{WQ2002} with model A and $\Lambda_{\mathrm{QCD}}=440$MeV, $m_\pi=139.6$MeV and $m=\frac{m^2_\pi}{2B_0}=12.3$MeV, substitute the resulting quark self-energy into the above formulae for the coefficients, and finally perform the numerical computations. Considering that some momentum integrations are divergent, we take $\Lambda=1$GeV as the cutoff parameter in the Euclidean space momentum integration. Expressing all results in units of $m_\pi$, values for the coefficients are listed in Table.\ref{coeff}
 
  \begin{table*}[h]
 \caption{\label{coeff}{The obtained coefficients in unit of $10^{-2}$.}}
 \vspace*{-0.5cm}\begin{eqnarray}
 \begin{array}{cccccccccccc}
 \hline\hline
 \displaystyle\frac{\phi_{\delta0}}{m^2_\pi}&\displaystyle\frac{\phi_{\delta1}}{m^2_\pi}&\displaystyle\frac{\phi_0}{m^2_\pi}&\displaystyle\frac{\phi_{10}}{m^2_\pi}&
 \displaystyle\frac{\phi_{11}}{m^2_\pi}&\displaystyle\frac{\phi_{p,\delta0}}{m^3_\pi}
 &\displaystyle\frac{\phi_{p,\delta10}}{m^3_\pi}  &\displaystyle\frac{\phi_{p,\delta11}}{m^3_\pi}&\displaystyle\frac{\phi_{p1}}{m^3_\pi}&\displaystyle \frac{\phi_{\sigma,0}}{m^2_\pi}&
\displaystyle\frac{\phi_{\sigma,10}}{m^2_\pi} &\displaystyle \frac{\phi_{\sigma,11}}{m^2_\pi}\\
 \hline
1.85&-0.15&10.13&0.98&-2.75&82.22&5.49&-0.16&0.86&8.13&-0.72&-0.17\\
 \hline\hline
 \end{array}\notag
 \end{eqnarray}
 \end{table*}
\noindent We see that the first-order corrections are orders of magnitude smaller than the leading order result. This verifies our conjecture that chiral perturbation will work well.\vspace*{0.5cm}~

To summarize, we have shown the GND model is a model which can correctly generate the minimal WT vertices. For PDAs, GND's result is similar to that of the NJL-like models, i.e. in the chiral limit except some end point pillars, PDAs take flat-like forms and non-flat effects of PDAs are due to nonzero pion and light-quark current mass corrections. We have shown that the variable $u$ in PDAs is just the Feynman parameter of loop calculations of standard perturbation theory and chiral perturbation works well for PDAs enabling quantitative estimates of PDA values. These results are valid not only for the GND model, but also for the larger class of chiral quark and NJL-like models.
Further, the leading order nonzero pion and current quark masses corrections lead concave type asymptotic-like form modifications for $\phi(u)$ with end-point pillars and $\phi_\sigma(u)$ above the flat-like form backgrounds as a substitution of original double-hump structure of CZ-like form of PDAs. To force matching our results with other phenomenological ones especially the end-point behaviors, one can take present GND model chiral limit result as a starting point, ignore present corrections from pion and current quark masses, perform the QCD Gegenbauer evolution for PDAs as done in Ref.\cite{flat}.
\section*{ACKNOWLEDGMENTS}
This work is supported by the National Science Foundation of China (NSFC) under Grants No.11075085, Specialized Research Fund Grants No.20110002110010 for the Doctoral Program of High Education of China, and Tsinghua University Initiative Scientific Research Program.



\begin{thebibliography}{1}
 \bibitem{GND}H.Yang, Q.Wang, Q.Lu, Phys. Lett. {\bf B532}, 240(2002).
 \bibitem{WQ2002}H.Yang, Q.Wang, Y.P.Kuang and Q.Lu, Phys. Rev. {\bf D66}, 014019(2002).
 \bibitem{WQ2010}S.Z.Jiang, Y.Zhang, C.Li, and Q.Wang, Phys. Rev. {\bf D81}, 014001(2010).
 \bibitem{WQseries}Y.L.Ma and Q.Wang, Phys. Lett. {\bf B560}, 188(2003);\\
 S.Z.Jiang and Q.Wang, Phys. Rev. {\bf D81}, 094937(2010);\\
 S.Z.Jiang, Y.Zhang, Q.Wang, arXiv:1203.0712v2[hep-ph].
 \bibitem{GS}J. Gasser and H. Leutwyler, Ann. Phys. {\bf 158}, 142(1984); Nucl. Phys. {\bf B250}, 465(1985).
  \bibitem{PS}H. Pagels and S. Stokar, Phys. Rev. {\bf D20}, 2947(1979).
  \bibitem{Rvertex}J.S.Ball and T.W.Chiu, Phys. Rev. {\bf D22}, 2542(1980); {\bf D23}, 3085(1981).
  \bibitem{Rvertex1}  M.R.Frank, K.L.Mitchell, C.D.Roberts, and P.C.Tandy, Phys. Lett. {\bf B359}, 17(1995);\\
  A.E.Dorokhov, W.Broniowski, and E. Ruiz Arriola, Phys. Rev. {\bf D74}, 054023(2006).
  \bibitem{GNC}B.Holdom, Phys. Rev. {\bf D45}, 2534(1992).
  \bibitem{VectorVertex}A.K{\i}z{\i}ers\"{u} and M.R.Pennington, Phys. Rev. {\bf D79},125020(2009);\\
   A.Bashir, R.Bermudez, L.Chang and C.Roberts, Phys.Rev. {\bf C85}, 045205(2012).
 \bibitem{Holdom}B.Holdom, R.Lewis, Phys. Rev. {\bf D51}, 6318(1995).
 \bibitem{asymp}G.P. Lepage and S.J. Brodsky, Phys. Rev. {\bf D22}, 2157(1980).
 \bibitem{CZ}V.L.Chernyak and A.R.Zhitnitsky, Nucl. Phys. {\bf B201}, 492(1982).
 \bibitem{flat}E.R. Arriola and W. Broniowski, Phys. Rev. {\bf D66}, 094016(2002).
\bibitem{WXG}X.G.Wu, T.Huang, T.Zhong, Chin. Phys. {\bf C37}, 063105(2013);\\
X.G.Wu, T.Huang, Phys. Rev. {\bf D84}, 074011(2011).
 \bibitem{nonlocalPDAs}P.Kotko and M.Praszalowicz, Phys. Rev. {\bf D81}, 034019(2010).
\bibitem{nonlocalPDAs1}M.Praszalowicz and A.Rostworowski, {\bf D64}, 074003(2001).
 \bibitem{instanton}D.Diakonov and V.Y.Petrov, Nucl. Phys. {\bf B245}, 259(1984); {\bf B272}, 457(1986);\\
 M.Praszalowicz and A.Rostworowski, Phys. Rev. {\bf D64}, 074003(2001).
 \bibitem{anlycity}P. Maris, Phys. Rev. {\bf D50}, 4189(1994); V.Sauli, Few Body Syst. {\bf 39}, 45(2006).
  \bibitem{DAdef}M.Beneke, Th.Feldmann,  Nucl.Phys. {\bf B592}, 3(2001).
 \end{thebibliography}
\end{document}